\documentstyle[12pt,aasms4]{article}

\received{3 February 1996}
\accepted{11 April 1996}
 
\lefthead{Bandiera et al.}
\righthead{SNR G11.2--0.3 and its Central Pulsar}
 
\outer\def\beginsection#1\par{\vskip0pt plus2\baselineskip\penalty-250
  \vskip0pt plus-2\baselineskip\bigskip\vskip\parskip
  \centerline{#1}\nobreak\smallskip}
 
\def\ov{\over} \def\({\left(} \def\){\right)} \def\E#1{\times10^{#1}}
\def\U#1{{\,\rm #1}}
 \def\ergs{erg\,s^{-1}} \def\cmc{cm^{-3}}
\def\D{\hbox{\it d}}
\def\td#1#2{{\mathchoice{\D#1\ov\D#2}{\D#1/\D#2}{\D#1/\D#2}{\D#1/\D#2}}}

\def\Edot{\dot{\cal E}} \def\al{\alpha} \def\tot{\tilde t}
\def\gm{\gamma}  \def\numax{\nu_{max}}
\def\eps{\epsilon}
\begin{document}
\title{The Supernova Remnant G11.2--0.3 and its central Pulsar}
\author{R. Bandiera}
\affil{Osservatorio Astrofisico di Arcetri, Largo E. Fermi 5, 50125
       Firenze, Italy}
\author{F. Pacini}
\affil{Osservatorio Astrofisico di Arcetri, and\\
       Dipartimento di Astronomia e Scienza dello Spazio,
       Universit\`a degli Studi,\\
       Largo E. Fermi 5, 50125 Firenze, Italy}
\and
\author{M. Salvati}
\affil{Osservatorio Astrofisico di Arcetri, Largo E. Fermi 5, 50125
       Firenze, Italy}
\begin{abstract}
The plerion inside the composite Supernova Remnant G11.2--0.3 appears to be
dominated by the magnetic field to an extent unprecedented among well known
cases. We discuss its evolution as determined by a central pulsar and the
interaction with the surrounding thermal remnant, which in turn interacts
with the ambient medium. We find that a plausible scenario exists, where
all the observations can be reproduced with rather typical values for the
parameters of the system; we also obtain the most likely period for the still 
undetected pulsar.
\end{abstract}
\section{Introduction}

Vasisht et al.\ (1996, hereafter VA) have recently discussed the nature of the
remnant G11.2--0.3 (Historical Supernova 386 AD) and have shown that it is
likely to be of a composite nature, as already suggested on the basis of radio
data by Morsi \& Reich (1987).
The hard X-ray emission observed by ASCA, combined with previous Einstein data
and observations at other wavelengths, strongly suggests (despite the fact that
a pulsar has not yet been discovered) the presence of a plerion surrounded by a
thermal shell.
The data across the electromagnetic spectrum can then be combined with the
theory of the evolution of pulsar-powered remnants (Pacini \& Salvati 1973
(PS); Bandiera, Pacini \& Salvati 1984; Reynold \& Chevalier 1984)
and allow a determination of the physical parameters of the system.
The main outcome of the interpretation proposed by VA is 
that the plerion should be
characterized by a magnetic energy much higher than the energy channelled into
relativistic electrons, at variance with, e.g., the Crab Nebula.

In our paper we study the evolution of the plerion coupled with that
of the surrounding shell.
We show that the magnetic field determination of VA is only marginally
consistent, from a dynamical point of view,
if the thermal and non-thermal remnants are in pressure equilibrium.
We identify, on the other hand, a scenario that is in agreement with the
observations, and derive the initial and present parameters of the central
pulsar.
In particular, we show that the apparent preponderance of the magnetic 
component in the pulsar output may be a natural outcome of the evolution, 
and does not imply a strong unbalance in the mechanism of production.

Following VA we assume that G11.2--0.3 is the remnant of SN 386 AD and
therefore has an age $t=1610\U{yr}=5\E{10}\U{s}$.
For its distance we use a reference value of $5\U{kpc}$ and define
$d_5=d/(5\U{kpc})$.
The outer thermal shell has radius $R_s=3.3\,d_5\U{pc}$, 
electron temperature $T_s=
0.73\U{keV}=8.5\E{6}\U{K}$, X-ray luminosity $L_s\sim10^{36}d_5^2\U{\ergs}$ (in
the 0.6--3.3 keV band), from which VA derive an ambient density $n_o\sim1.5
\,d_5^{-1/2}\U{\cmc}$, by assuming a Sedov (1959) expansion in the 
interstellar medium.
The central plerionic component has a radius $R_p\sim1\,d_5\U{pc}$ and its
spectrum is equal to
\begin{equation}
L_{\nu}\simeq3\E{16}\(\nu\ov2.5\E{18}\U{Hz}\)^{-0.8} d_5^2\U{\ergs Hz^{-1}},
\eqnum{1}
\end{equation}
corresponding to an X-ray luminosity $L_p\simeq1\E{35}d_5^2\U{\ergs}$, in the
0.5--4 keV band.
eq.~(1) is consistent with the observations down to 32 GHz, but overestimates
the plerion emission at 1 GHz.
Accordingly, the spectrum must have a break at a frequency $\nu_o$ 
between 1~GHz and 32~GHz.

\section{The model}

The assumption of the PS model is that the central pulsar feeds both magnetic
field and relativistic particles into the expanding plerion.
Taking into account the adiabatic losses and the rate of production, the
evolution of the field strength is given by
\begin{equation}
\td{}t\(B^2R_p^4\ov6\)=p\Edot R_p.
\eqnum{2}
\end{equation}

Here $p$ is the fraction of the pulsar energy loss $\Edot$ that goes into
magnetic energy; it is assumed that $p$ is constant.
According to standard pulsar electrodynamics $\dot\Omega\propto-\Omega^n$,
with $n$ being the so-called braking index, and
\begin{equation}
\Edot={\Edot_o\ov\(1+t/\tau_o\)^\al},
\eqnum{3}
\end{equation}
with $\al=(n+1)/(n-1)=2$ in a pure dipole field. Measured values of $n$ range
roughly between 1.4 for the Vela pulsar (Lyne, private communication) and 2.8
for PSR 1509-58 (Kaspi et al. 1994).
The pulsar is assumed to inject also relativistic particles, whose energies are
distributed according to a power law with constant index $\gm$, at a rate
$(1-p)\Edot$.
The evolution of the electron energies is determined by both adiabatic and
synchrotron losses.

An immediate implication of the theory is that the emitted spectrum of the
plerion should show a break at a frequency $\nu_b$ such that particles
radiating at $\nu_b$ have a lifetime equal to the age of the nebula.
Since $\nu_b\propto B^{-3}t^{-2}$, if the observed break is identified with
$\nu_b$ one directly derives the field strength in G11.2--0.3
\begin{equation}
B=2.1\E{-3}\(\nu_b\ov32\U{GHz}\)^{-1/3}\U{G},
\eqnum{4}
\end{equation}
which corresponds to a total field energy
\begin{equation}
W_B=2.1\E{49}\(\nu_b\ov32\U{GHz}\)^{-2/3}d_5^3\U{erg}.
\eqnum{5}
\end{equation}

The synchrotron spectrum above $\nu_b$, up to a maximum frequency $\numax$, is
given by
\begin{equation}
L_{\nu}={2-\gm\ov\gm-1}{(1-p)\Edot\ov2\numax}\(\nu\ov\numax\)^{-\gm/2}.
\eqnum{6}
\end{equation}

From the observations $\gm=1.6$ and $\numax>2.5\E{18}\U{Hz}$; by combining
eqs.~(1) and (6) we find the present-time energy loss from the pulsar:
\begin{equation}
\Edot=2.24\E{35}\(\numax\ov2.5\E{18}\U{Hz}\)^{0.2}{d_5^2\ov1-p}\U{\ergs}.
\eqnum{7}
\end{equation}

We describe the relative importance of the magnetic and particle components
in terms of the dimensionless quantity $Q=B^2R_p^3/6(1-p)\Edot t$. 
By substituting eq.~(3) into eq.~(2) and integrating, we find
\begin{equation}
Q={p\ov1-p}{(1+\tot)^\al\ov\tot^{1+r}}\int_0^{\tot}{x^r\ov(1+x)^\al}dx;
\eqnum{8}
\end{equation}
here we put $\tot=t/\tau_o$, and assumed a power law for the expansion,
$R_p\propto t^r$. Note that $Q$ is a measure of the balance
at production rather than a gauge of equipartition: in fact,
when $p\approx0.5$, $Q\approx1$ until $t\le\tau_o$ (and later on as well
if $\al=1+r$). The ratio of magnetic to particle energy, instead, can
be arbitrarily high if the radiative lifetime of the energetically
important particles is sufficiently shorter than $t$. On the observational 
side, eqs.~(5) and (7) give
\begin{equation}
Q=1860\(\nu_b\ov32\U{GHz}\)^{-2/3}\(\numax\ov2.5\E{18}\U{Hz}\)^{-0.2}\!\!d_5:
\eqnum{9}
\end{equation}
the magnitude of $Q$ in the present case, to be compared with $Q\approx 1$ 
for the Crab plerion, is a quantitative measure of the discrepancy which
we want to address.
For G11.2--0.3, then, either $p$ is very close to unity ($1-p<10^{-3}$), 
or $\tot\gg 1$, $(\alpha-1-r)>0$.
We regard the former possibility as very unlikely, and in
the following we shall limit ourselves only the latter.

When $t\gg\tau_o$, it is appropriate to assume flux conservation for the 
evolution of the magnetic field, $B\propto t^{-2r}$; furthermore,
the plerion spectrum shows {\it two} breaks (PS), of which the higher one,
$\nu_b$,
corresponds to the radiative lifetime at the current time $t$, and
the lower one, $\nu_c$, is the signature of the radiative lifetime at $\tau_o$.
If we identify the observed break with $\nu_b$, eq.~(4) holds, and the
pressure interior to the plerion is relatively very high. 
As we show in the following,
the magnetic pressure corresponding to eq.~(4) exceeds the shell pressure
by a large factor; this is possible only if the plerion expansion
is limited by the inertia of some cold matter tied to the plerion itself,
in which case we expect $r\ge 1$. Then $r\ge 1$ and $\alpha\le 3$, as observed
in all measured cases, give $p$ of order 0.5 only at the expense 
of $\tot\ge 1000$,
and an initial magnetic energy of order $\tot\times W_B > 10^{52} d^3_5$ erg.
We consider very unlikely a value so much larger than the canonical supernova
output, and propose that: $i$) the plerion and the shell are in pressure
equilibrium, so that $r<1$; and $ii$) the observed break is to be identified
with $\nu_c$, so that $\nu_b=\nu_c\times\tot^{10r-2}\gg\nu_o$, 
and $Q$ is decreased with respect to eq.~(9).
However, the break at $\nu_b$ entails a change of slope
$\eps=(2.8r-0.4\al)/(5r-1)$ (see PS); since G11.2--0.3 shows a straight 
spectrum from the microwaves to the X~rays, we must require
$\eps=0,~~\al=7r.$

Within this framework more detailed results can be obtained by coupling
the plerion evolution to the hydrodynamics of the thermal shell.
We can distinguish two different cases, according to the distribution of the
ambient density being of the interstellar type ($\rho_o\sim$ constant, 
henceforth IS), or circumstellar type ($\rho_o\sim r^{-2}$, henceforth
CS). If the ambient medium has a constant density, $R_s\propto
t^{2/5}$, $R_p\propto t^{3/10}$, and, because of the condition
on $\eps$, $\al=2.1$.
If the shell expands inside the pre-supernova wind, one has $R_s\propto t^{2
/3}$, $R_p\propto t^{4/7}$, and $\al=4.0$.

We have computed in detail the two cases, and give the results in Table~1.
In particular, we have adjusted the unknown distance $d_5$ so as to make
the total energy of the shell remnant equal to the canonical 10$^{51}$ erg.
We have chosen the value of $\nu_o$ within the allowed interval so as
to bring our results as close as possible to `typical' values. 
And, finally, we have neglected the
work done by the plerion on the shell remnant, which is justified {\it a
posteriori} by the relatively small plerion energetics resulting from the
computations. The precise value of the ambient density in the IS and CS
cases was computed so as to reproduce the temperature and flux of the
thermal X~rays given in VA.

Note that the CS assumption leads to very `palatable' estimates for all the
unknowns, with one exception: the slowing down exponent $\al$ 
implies a braking index $n=1.67$. We note that such value is
still in the acceptable range between a purely dipolar ($\al=3,~~n=3$) and
a purely monopolar geometry ($\al=\infty,~~n=1$), but that for most pulsars
the braking index is between 2 and 3. On the other hand,
the IS hypothesis carries along some deviation from perfect balance at 
production ($p\sim0.9$) and a pulsar magnetic field at the extreme 
of the observed range ($B_*=1.3~10^{13}$~G, like in PSR~1509-58).
All in all, we regard the latter interpretation as the
most plausible one, also because it places G11.2--0.3 exactly on the empirical
relation between $L_p$ and $\Edot$ given by Seward \& Wang (1988); we
suggest that the corresponding values in Table~1 be taken as a reasonable
guess for further observing efforts.

A major improvement in the modelling could ensue if future observations
were to give the precise frequency of the break(s) in the spectrum
and the related change(s) of slope, or the detection of the pulsar.

\section{Conclusions}

If G11.2--03 is a composite remnant deriving from the historical explosion of
SN 386 AD, the main parameters of the nebula and of the central pulsar can be
estimated in the framework of the PS model.
They appear to be consistent with the properties of the remnant and the
standard views on pulsars.

The apparent preponderance of the magnetic channel in the output of
the central pulsar
is the consequence of the dynamical evolution under the influence of
the external shell.
This has slowed down the expansion and therefore reduced the importance of the
adiabatic losses with respect to the synchrotron losses.
Since most of the energy was released by the pulsar during a relatively short
initial phase, it is natural that the subsequent evolution has led to a large 
value of $Q$, eq.~(8), even if the pulsar output was always well balanced, 
$p\approx0.5$.
We strongly stress this point as relevant to all sorts of non-thermal sources:
only if the generation of particles and fields keeps occurring for a time scale
longer or roughly equal to the age of the source does $Q$ reflect the balance at
production between fields and particles. This is why $Q\approx1$ in the 
Crab plerion,
for which $t\approx3\times\tau_o$. In addition, and for quite independent
reasons, the Crab is not far from equipartition,
since its $\nu_b$ is not far from the frequency where most of the power is
emitted.

\acknowledgments
 
This work was partly supported by a Grant of the Italian Space Agency.

\begin{deluxetable}{lrr}
\tablenum{1}
\tablecaption{Parameters of the Models}
\tablewidth{0pt}
\tablehead{\colhead{Parameter}&\colhead{IS Model}&\colhead{CS Model}}
\startdata
$\nu_o     $&$4\U{GHz}           $&$32\U{GHz}          $\nl
$d         $&$9.2\U{kpc}         $&$7.4\U{kpc}         $\nl
$p         $&$0.9                $&$0.3                $\nl
$B_p       $&$9.3\E{-4}\U{G}     $&$1.8\E{-4}\U{G}     $\nl
$P         $&$0.34\U{s}          $&$2.73\U{s}          $\nl
$\Edot     $&$1.55\E{37}\U{\ergs}$&$1.25\E{36}\U{\ergs}$\nl
$B_*       $&$2.5\E{13}\U{G}     $&$1.6\E{12}\U{G}     $\nl
$\tau_o    $&$5.8\E8\U{s}        $&$7.0\E9\U{s}        $\nl
$P_o       $&$0.029\U{s}         $&$0.139\U{s}         $\nl
${\cal E}_o$&$2.4\E{49}\U{erg}   $&$1.0\E{48}\U{erg}   $\nl
\enddata
\end{deluxetable}

\clearpage

\end{document}